\documentclass[preprint]{JHEP3} 
\JHEPspecialurl{http://jhep.sissa.it/JOURNAL/JHEP3.tar.gz}
\usepackage{epsfig,multicol}
\newcommand\fverb{\setbox\fverbbox=\hbox\bgroup\verb}
\newcommand\fverbdo{\egroup\medskip\noindent%
            \fbox{\unhbox\fverbbox}\ }
\newcommand\fverbit{\egroup\item[\fbox{\unhbox\fverbbox}]}
\newbox\fverbbox

\newcommand{\be}{\begin{equation}}
\newcommand{\dd}{\displaystyle}
\newcommand{\ee}{\end{equation}}
\newcommand{\ba}{\overline }
\newcommand{\bref}[1]{(\ref{#1})}

\newcommand{\bea}{\begin{eqnarray}}
\newcommand{\eea}{\end{eqnarray}}

\def\pa{\partial}
\def\7{\tilde}
\def\8{\hat}

\newcommand{\nn}{\nonumber}

 \def\slash#1{\setbox0=\hbox{$#1$}#1\hskip-\wd0\dimen0=5pt\advance
       \dimen0 by-\ht0\advance\dimen0 by\dp0\lower0.5\dimen0\hbox
         to\wd0{\hss\sl/\/\hss}}
\def\l{{\ell}}

\def\v{\kappa}\def\alp{{\mu}}
\def\balpha{{\alpha\hskip-2.2mm\alpha}}\def\bbeta{{\beta\hskip-2.1mm\beta}}

\title{
Space-time Vector Supersymmetry and Massive Spinning Particle}

\author{{Roberto Casalbuoni}\\
    {Department of Physics, University of Florence,
         INFN-Florence and Galileo Galilei Institute, Florence\, Italy}\\
    E-mail: \email{casalbuoni@fi.infn.it}}
\author{{Joaquim Gomis}\\
    {PH-TH Division, CERN, 1211 Geneva 23, Switzerland\\
 Departament d'Estructura i Constituents de la
Mat\`eria and \\
Institut de Ciencies del Cosmos, Universitat de Barcelona,
 Diagonal 647, 08028
Barcelona,
Spain}\\
    E-mail: \email{gomis@ecm.ub.es}}
\author{{Kiyoshi Kamimura}\\
    {Department of Physics, Toho University,
Funabashi, 274-8510 Japan}\\
    E-mail: \email{kamimura@ph.sci.toho-u.ac.jp}}
\author{{Giorgio Longhi}\\
    {Department of Physics,
University of Florence, and
INFN-Florence, Italy}\\
    E-mail: \email{longhi@fi.infn.it}}

\received{\today}       
\accepted{\today}       

\preprint{{PH-TH CERN/2007/243, UB-ECM-PF-07-36, Toho-CP-0885 }}

\abstract{

We construct the action of a relativistic spinning particle from a
non-linear realization of a space-time odd vector extension of the
Poincar\'e group. For particular values of the parameters appearing
in the lagrangian the model has a gauge world-line supersymmetry.{
As a consequence of this local symmetry there are BPS solutions in
the model preserving 1/5 of the supersymmetries.} A supersymmetric
invariant  quantization produces two decoupled 4d Dirac equations. }


\begin{document}


\section{Introduction \label{sec:0}}
Supersymmetry plays a crucial role in field theories, supergravities
and String/M theory. In flat space-time supersymmetry is
characterized by the presence of odd spinor charges that together
with the generators of the Poincar\'e group form the target space
Super Poincar\'e group.

Strings with the space-time supersymmetry are described by the
Green-Schwarz action \cite{Green:1983wt}. The fermionic components
of the string are realized by spinorial fields in target space. The
associated particle  model (superparticle), that was introduced
before \cite{Casalbuoni:1976tz,Brink:1981nb} has, for a particular
value of the two coefficients of the lagrangian, i.e., for the
parameters of Nambu-Goto  and Wess-Zumino pieces, a fermionic gauge
symmetry called kappa symmetry \cite{deAzcarraga:1982dw}
\cite{Siegel:1983hh}. The covariant quantization of this model is an
unsolved issue.

There is an alternative to the GS string known as spinning string,
or Neveu-Schwarz-Ramond string
\cite{Deser:1976rb}\footnote{formulated prior to the GS string},
that has world "line" supersymmetry. The fermionic components of the
string are described by odd vector fields in target space. A
truncation of this theory known as GSO projection
\cite{Gliozzi:1976qd} produces a spectrum that is space-time
supersymmetric invariant.

The corresponding particle model (spinning particle) was introduced
in refs. \cite{Barducci:1976qu,Brink:1976sz,Berezin:1976eg}. The
quantization of this model reproduces the four dimensional Dirac
equation.

In this paper we will consider an odd vector extension of the
Poincar\'e group, to be called the Vector Super Poincar\'e group G,
first formulated in  \cite{Barducci:1976qu}. This symmetry was
introduced  with the aim of obtaining a pseudo-classical description
of the Dirac equation. However this result was obtained by using a
constraint breaking the symmetry itself. In this paper we want to
take full advantage of this new symmetry. We will show that the
massive spinning particle action of reference \cite{Barducci:1976qu}
can be obtained by applying the method of non-linear realizations
\cite{Coleman} to this group. However, trying to preserve the target
space supersymmetry under quantization we will obtain two copies of
the 4d Dirac equation. On the other hand, breaking the rigid
supersymmetry by a suitable constraint on the Grassmann variables we
will recover the 4d Dirac equation of
\cite{Barducci:1976qu,Brink:1976sz,Berezin:1976eg}.

The lagrangian will contain a Dirac-Nambu-Goto piece and two
Wess-Zumino terms. By construction the action is invariant under
rigid vector supersymmetry.
 For particular values of the coefficients of the lagrangian,
the model has world line gauge supersymmetry which is analogous to
the fermionic kappa symmetry of the superparticle case. When we
require world-line supersymmetry the model has bosonic BPS
configurations that preserve 1/5 of the vector supersymmetry. The
BPS configurations imply second order equations of motion.

The organization of the paper is as follows: In section 2 we will
introduce the space-time vector supersymmetry. In section 3 we will
construct the massive spinning particle action using the method of
non-linear realizations. Section 4 is devoted to the canonical
formalism. In section 5 we will quantize the model in terms of the
unconstrained variables. In section 6 we will quantize the model in
the Clifford representation. Section 7 will be devoted to classical
BPS configurations and finally in section 8 we will give some
conclusions.

\section{Space-time Vector Supersymmetry}
Let us consider the Poincar\'e algebra\footnote{ We consider a four
dimensional space-time with metric
${\rm{diag}}\,\eta_{\mu\nu}=(-;+++).$} with generators
$P_\mu,M_{\mu\nu}$ extended with odd graded generators,
supertranslations, belonging to a pseudoscalar $(G_5)$  and a pseudovector
$(G_\mu)$ representation of the Lorentz group  and two central
charges $Z$ and $\tilde Z$. They satisfy 
\be [M_{\mu\nu},M_{\rho\sigma}]=-i\eta_{\nu\rho}M_{\mu\sigma}-
i\eta_{\mu\sigma}M_{\nu\rho}+i\eta_{\nu\sigma}M_{\mu\rho}+
i\eta_{\mu\rho}M_{\nu\sigma},\label{eq:ext0} \ee \be
[M_{\mu\nu},P_\rho]=i\eta_{\mu\rho}P_\nu-i\eta_{\nu\rho}P_\mu,\qquad
[M_{\mu\nu},G_\rho]=i\eta_{\mu\rho}G_\nu-i\eta_{\nu\rho}G_\mu,
\label{eq:ext01} \ee \be [G_\mu,G_\nu ]_+=\eta_{\mu\nu}Z,~~~[G_5,G_5
]_+=\tilde Z \label{eq:ext1},\ee \be
[G_\mu,G_5]_+=-P_\mu\label{eq:ext2}. \ee We first consider a coset
of the vector Super Poincar\'e group, $G/O(3,1)$\footnote{Note that
the algebra of vector supersymmetry is a subalgebra of the N=2
topological supersymmetry algebra in four dimensions. The
topological algebra contains also an odd  self dual tensor
generator, see for example \cite{Alvarez:1994ii} and the references
there. } vector of supersymmetry , and parameterize the group
element as \be g=e^{iP_\mu x^\mu} e^{iG_5\xi^5}e^{i G_\mu
\xi^\mu}e^{iZc} e^{i\tilde Z\tilde c}.\label{eq:coset}\ee
 $x^\mu,\;\xi^\mu $ and $ \xi^5$  are the superspace coordinates and
 $c$ and $\tilde c$  are coordinates associated to the
projective representations of G. This is analogous to the ordinary
case of the N=2 Super Poincar\'e group with central charges. Notice
that we assume the odd generators $G_\mu$ and $G_5$  anticommute
with the Grassmann coset parameters $\xi^\mu$ and $\xi^5$. It
follows that the representation of the group is unitary if these
generators are anti-hermitian.

The MC 1-form associated to this coset is \bea \Omega=-ig^{-1}dg
&=&P_\mu \left(dx^\mu- i\xi^\mu d\xi^5\right)+G_5 d\xi^5 +G_\mu
d\xi^\mu+\nn\\&+&Z\left(dc-\frac i 2d\xi_\mu \xi^\mu\right)+\tilde
Z\left(d\tilde c-\frac i 2d\xi^5\xi^5\right). \eea From this we get
the even differential 1-forms\be L_x^\mu=dx^\mu- i \xi^\mu
d\xi^5,~~~L_Z=dc+\frac i 2 \xi^\mu d\xi_\mu,~~~L_{\tilde Z}=d\tilde
c+\frac i 2\xi^5 d\xi^5 \label{1formb}\ee and the odd ones by \be
L_\xi^\mu=d\xi^\mu,~~~L_\xi^5=d\xi^5. \label{1formf}\ee

The supersymmetry transformations leaving the MC form invariant are
\be \delta x^\mu= i \epsilon^\mu
\xi^5,~~~\delta\xi^\mu=\epsilon^\mu,~~~\delta c=\frac i
2\xi_\mu\epsilon^\mu\ee and \be
\delta\xi^5=\epsilon^5,~~~\delta\tilde c=\frac i2\xi^5\epsilon^5.\ee
These vector supersymmetry transformations are generated by $G^\mu$
and $G^5$ respectively and were first discussed in
\cite{Barducci:1976qu}. The transformations generated by $P_\mu,Z,
\tilde{Z}$ are
the infinitesimal translations of 
the coordinates $x^\mu, c$ and $\tilde c$,
\be \delta
x^\mu= a^\mu,\qquad \delta c=\epsilon_Z,\qquad
\delta\tilde{c}=\epsilon_{\tilde{Z}}. \ee
The vector fields generating the previous transformations are
\bea
X_\mu^G&=&-i\left(\frac{\partial}{\partial \xi^\mu}+
i\xi^5\frac{\partial}{\partial x^\mu}- \frac i2\xi_\mu
\frac{\partial}{\partial c}\right), \qquad
X_5^G=-i\left(\frac{\partial}{\partial\xi^5}-\frac i2\xi^5
\frac{\partial}{\partial\tilde{c}}\right) ,
\nn\\
X_\mu^P&=&-i\frac{\partial}{\partial x^\mu}, \qquad
X_Z=-i\frac{\partial}{\partial c}, \qquad
X_{\tilde{Z}}=-i\frac{\partial}{\partial\tilde{c}}.
\eea The algebra of
these vector fields\footnote{ Remember
$\frac{\partial}{\partial\xi^\mu}$ and
$\frac{\partial}{\partial\xi^5} $ are hermitian and $X_\mu^G$ and
$X_\mu^5$ are anti-hermitian. } is \be
[X_\mu^G,X_\nu^G]_+=-\eta_{\mu\nu}X_Z,  \qquad
[X_5^G,X_5^G]_+=-X_{\tilde{Z}},
 \qquad [X_\mu^G,X_5^G]_+=X_\mu^P. \ee
With the Lorentz generators $X^M_{\mu\nu}$, these vector fields give
a realization of the Vector Super Poincar\'e algebra.\footnote{The
reason for the overall sign difference from the starting algebra  is
that now the generators are {\it active} operators.}  The
representations of this algebra will be studied in their general
aspects in \cite{nuovo}.

\section{Massive Spinning Particle Lagrangian}
In this section we will study how to construct the massive spinning
particle action from a non-linear realization \cite{Coleman}  of the
Vector Super Poincar\'e group G. The relevant coset is G/O$(3)$
defined by the little group of a massive particle 
O(3) as the unbroken (stability) group of the coset. We write the
elements of the coset as \be g=g_L\; U,~~~U=e^{i M_{0i}v^i},\ee
where $g_L$ is the same as in  \bref{eq:coset} \be g_L=e^{iP_\mu
x^\mu} e^{iG_5\xi^5}e^{i G_\mu \xi^\mu}e^{iZc} e^{i\tilde Z\tilde
c}\label{eq:coset2}\ee
 and $U$ 
represents a finite Lorentz boost with   parameters $v^i$.

The Maurer-Cartan 1-form is now \be \Omega=-i g^{-1}dg
=U^{-1}\Omega_L U-i U^{-1} dU,\qquad \Omega_L\equiv
-ig_L^{-1}dg_L.\ee Using the commutation relations of the Lorentz
generators  with the four-vectors $P_\mu$ and $G_\mu$
\bref{eq:ext01} we find \bea U^{-1}P_\mu U&=&P_\nu
{\Lambda}^\nu_{\cdot\mu}(v),\qquad U^{-1}G_\mu U=G_\nu
{\Lambda}^\nu_{\cdot\mu}(v), \eea where
${\Lambda}^\nu_{\cdot\mu}(v)$ is a finite Lorentz boost \bea
{\Lambda}^\nu_{\cdot\mu}=\pmatrix{\cosh v&-v^i\frac{\sinh v}{v}\cr
-\frac{\sinh v}{v}v_j&\delta^i_j+\frac{v^iv_j}{v^2}(\cosh
v-1)},\qquad~~~ v\equiv\sqrt{(v^i)^2}.
 \label{Lambdaboost}\eea
It follows 
\be \Omega=P_\nu {\tilde L}^\nu_x+G_\nu {\tilde L}^\nu_\xi+G_5
L^5_\xi+Z L_Z+\tilde Z L_{\tilde
Z}+M_{0i}L^i_v+\frac12M_{ij}L^{ij}_v, \ee where
\be {\tilde
L}^\nu_x=\Lambda^\nu_{\cdot\mu}(v)L^\mu_x,~~~~ {\tilde L}^\nu_\xi=
\Lambda^\nu_{\cdot\mu}(v)L^\mu_\xi\ee
and the 1-forms $L$'s are given in (\ref{1formb}) and  (\ref{1formf}).
$L_v^i$ and $L_v^{ij}$ are given by \bea
L_v^i&=&dv^i+dv^j\left({\delta_j}^i-\frac{v^jv^i}{v^2}\right)\left(\frac{\sinh
v}v-1\right),\nn\\
L^{ij}&=&\frac{{dv^{i}v^{j}}-{dv^{j}v^{i}}}{v^2}(\cosh v-1).\eea

The invariant action for the particle is  a sum of the manifest
$O(3)$ invariant one forms. By taking the pull-back, the (Goldstone)
super-coordinates become functions of the parameter $\tau$ that
parameterizes the worldline of the particle, see for example
\cite{Gomis:2006xw}. The action is \be\label{actionv}
S[x(\tau),\xi(\tau),v(\tau)]=\int \left(-{\alp}\,\tilde L^0 _x-\beta
L_Z-\gamma L_{\tilde Z}\right)^*=\int L\;d\tau, \ee where * means
pull-back on the world line and ${\alp}, \beta$ and $\gamma$ are
real constants to be identified with the mass and the central
charges, $ m,\, Z$ and $\tilde Z$ respectively. The action is
invariant under global Vector Super Poincar\'e transformations.
In addition it is invariant 
under a local supersymmetry transformation if the parameters satisfy
\be\label{condition1}-{\beta\gamma}={{\alp}^2}.
\label{BPScond}\ee In this paper we will study the case in which
this condition is satisfied. If we use the explicit form of the
finite Lorentz boost \bref{Lambdaboost} the action is
\be
S[x(\tau),\xi(\tau),v(\tau)]=\int\left[\,-{\alp}\left( L^0_x\cosh
v-L^i_x v^i\frac{\sinh v} v\right)-\beta L_Z-\gamma L_{\tilde
Z}\right]^*.\label{eq:65}\ee The explicit form of the local
supersymmetry transformation is \be
\delta\xi^0=-\frac{{\alp}}{\beta}\cosh
v\;\kappa^5(\tau),~~~\delta\xi^i=-\frac{{\alp}}{\beta}\frac{v^i}v\sinh
v \;\kappa^5(\tau),~~~\delta\xi^5=\kappa^5(\tau) \label{eq:77}\ee
and \be  \delta x^\mu= i
 \xi^\mu \v^5(\tau),~~~~\delta c=-\frac i 2\xi_\mu \delta \xi^\mu, ~~~\delta\tilde
c=-\frac i 2 \xi^5 \v^5(\tau),~~~~\delta v^a=0.\label{eq:351}\ee

 Now we will see how one eliminates the boost parameters $v^i$ from the theory.
This can be done  by using their equations of motion \be
\frac{\delta S}{\delta v^i}=0=L^0_{\tau}\frac {v^i}v\sinh
v{-}L_{\tau}^j\left[\delta^{ij}\frac{\sinh v}v +\frac
{v^iv^j}{v^2}\left(\cosh v- \frac{\sinh v}v\right)\right]
\label{eq69a}.\ee By solving this equation we get \be
\frac{v^i}{v}=\frac{L_{\tau}^i}{\sqrt{(L_{\tau}^j)^2}},~~~~ \sinh
v=\frac{\sqrt{(L_{\tau}^j)^2}}{\sqrt{(L_{\tau}^0)^2-(L_{\tau}^j)^2}}
,\qquad \cosh
v=\frac{L_{\tau}^0}{\sqrt{(L_{\tau}^0)^2-(L_{\tau}^j)^2}},
\label{eq:3.15}\ee where $L^\mu_\tau$ is the pull-back of the
$L_x^\mu$ \be L_\tau^\mu={\dot x}^\mu-i\xi^\mu{\dot\xi}_5.\ee Using
them in \bref{eq:65} we get \be S\left[x(\tau),\xi(\tau)\right]=\int
d\tau\left(-{\alp}\sqrt{-\left({\dot x}^\mu- i
\xi^\mu{\dot\xi}^5\right)^2}-\beta\left(\dot c +\frac i 2\xi^\mu
\dot\xi_\mu\right)-\gamma\left(\dot{\tilde c}+\frac i2\xi^5
\dot{\xi^5}\right)\right).\label{eq:652}\ee The local super
transformation for the action \bref{eq:652} is \be
\delta\xi^\mu=-\frac{p^\mu}{\beta}
\v^5(\tau),~~~\delta\xi^5=\v^5(\tau), \ee where $p_\mu$ is the
momentum conjugate to $x^\mu$ whereas the transformations for the
other variables remain the same as in \bref{eq:351}. Note that this
transformation is analogous to the kappa symmetry transformation of
the massive superparticle action \cite{deAzcarraga:1982dw}. In the
literature it is known as gauge world-line supersymmetry.

The action \bref{eq:652} coincides with the action originally
proposed in \cite{Barducci:1976qu} after we make a suitable
rescaling of the coordinates \be
\xi^{'\mu}=\sqrt{|\beta|}\,\xi^\mu,~~~\xi^{'5}=\sqrt{\gamma}\,\xi^5,~~~
c^{\,\prime}=|\beta|c,~~~{\tilde c}^{\,'}=\gamma \tilde
c.\label{eq:51}\ee Correspondingly the local super transformation
becomes \be \delta {x'}^\mu=\frac i \mu{\xi'}^\mu
{\kappa'}^5,~~~\delta{\xi'}^\mu=\frac
{p^\mu}{\mu}{\v'}^5(\tau),~~~\delta{\xi'}^5={\v'}^5(\tau), \ee
{where the parameter has also been rescaled as
${\kappa'}^5=\sqrt{\gamma} {\kappa}^5$ }.

\section{Canonical Formalism}

The canonical momentum conjugate to $x^\mu$ defined by the
lagrangian (\ref{actionv}) is \be  p_\mu=\frac{\partial L}{\partial
{\dot x}^\mu}=-{\alp} \Lambda_{\cdot\mu}^0(v).\ee Since
$\Lambda_{\cdot\mu}^0(v)$ is a time-like vector, the momentum
verifies the mass-shell constraint $p^2+{\alp}^2=0$. Therefore the
parameter ${\alp}$ is identified with the mass of the particle  and
the lagrangian \bref{actionv} can be written in the first order form
as
 \be L^C=p_\mu(\dot x^\mu-{i}\xi^\mu\dot\xi^5)
-\beta\frac{i}2\xi_\mu\dot\xi^\mu-\gamma\frac{i}2\xi^5\dot\xi^5
-\frac{e}2(p^2+{\alp}^2),\label{LagC}\ee where $e$ is a lagrange
multiplier (ein-bein). In the lagrangian we have omitted $\dot c$
and $\dot{\tilde c}$ terms since they are total derivatives. By
eliminating $p_\mu$'s and $e$ using their equations of motion, the
lagrangian goes back to the covariant one in eq. \bref{eq:652}.

Let us  study the constraints and symmetries of $L^C$. The
canonical momenta  are defined by using {\it left derivatives} as
the Grassmann variables are concerned, \bea
&&p^x_\mu=\frac{\partial L^C}{\partial \dot x^\mu}=p_\mu, \qquad
p^p_\mu=\frac{\partial L^C}{\partial \dot p^\mu}=0,  \qquad
p^e=\frac{\partial L^C}{\partial \dot e}=0, \nn\\
&&\pi_\mu=\frac{\partial^\l L^C}{\partial \dot \xi^\mu }=
\beta\frac{i}2\xi_\mu,   \qquad
\pi_5=\frac{\partial^\l L^C}{\partial \dot \xi^5 }=
\gamma\frac{i}2\xi^5+ip_\mu\xi^\mu. \eea All of them give rise to
primary constraints \bea
\phi^x_\mu&=&p^x_\mu-p_\mu=0,   \qquad
\phi^p_\mu=p^p_\mu=0,   \qquad
\phi^e=p^e=0, \nn\\
\chi_\mu&=&\pi_\mu-\beta\frac{i}2\xi_\mu=0,  \qquad
\chi_5=\pi_5-\gamma\frac{i}2\xi^5-ip_\mu\xi^\mu=0.
\label{priconst}\eea
The Hamiltonian, $H\equiv\dot q p-L$, is 
\bea H=\lambda_x^\mu \phi^x_\mu+\lambda_p^\mu \phi^p_\mu+\lambda_e
\phi^e +\lambda^\mu_\xi \chi_\mu+ \lambda^5_\xi
\chi_5+\frac{e}2(p^2+{\alp}^2), \eea where the $\lambda$'s are Dirac
multipliers. Using the graded Poisson brackets $\{p,q\}=-1$ we study
the stability of the constraints. We get the following secondary
constraint \be \phi\equiv\frac{1}2(p^2+{\alp}^2)=0,
\label{massshell}\ee and \be \lambda^p_\mu=0,\qquad
\lambda^x_\mu=ep_\mu+i\xi_\mu\lambda^5_\xi,\qquad
\lambda_\mu^\xi=-\frac{1}{\beta}\,p_\mu\lambda^5_\xi, \ee where we
have used the condition (\ref{BPScond}), $\beta\gamma=-{\alp}^2$.
The secondary constraint (\ref{massshell}) is preserved in time 
\be \dot\phi=0, \ee and it does not generate further constraints.
$\phi^x_\mu=\phi^p_\mu=0$ are  second class constraints and are used
to eliminate $p^x_\mu$ and $p^p_\mu.$  The second class constraints
$\chi_\mu=0$ are used to eliminate $\pi_\mu$. The Dirac bracket for
the remaining variables are \be
\{p_\mu,x^\nu\}^*=-{\delta_\mu}^\nu,\qquad
\{\xi^\mu,\xi^\nu\}^*=\frac{i}\beta{\eta^{\mu\nu}},\qquad
\{\pi_5,\xi^5\}^*=-1 \label{DB}\ee and the Hamiltonian becomes \bea
H=\lambda_e \phi^e+{e}\phi+\lambda^5_\xi \chi_5= \lambda_e
p^e+\frac{e}2(p^2+{\alp}^2)+\lambda^5_\xi(\pi_5-\gamma\frac{i}2\xi^5-ip_\mu\xi^\mu).
\eea The constraints $ \phi^e, \phi$ and $\chi_5$ appearing here are
the first class constraints. In particular we have\be
\{\chi_5,\chi_5\}^*=
i\gamma-\frac{i}{\beta}p_\mu p_\nu\eta^{\mu\nu}=-\frac{2i}{\beta}\;\phi.
\ee
 $\chi_5$ generates the local kappa variation corresponding to
(\ref{eq:77}),
 \be \delta x^\mu= i \xi^\mu
\v^5(\tau),~~~\delta\xi^\mu=-\frac{p^\mu}{\beta}
\v^5(\tau),~~~\delta\xi^5=\v^5(\tau),~~~ \delta
e=-\frac{2i}{\beta}\,\dot\xi^5\v^5(\tau),\label{kappavari2}\ee under
which the lagrangian transforms as \be \delta
L^C=\frac{d}{d\tau}\left(\frac i 2(p_\mu\xi^\mu-
\gamma{\xi^5})\v^5(\tau)\right). \ee The global vector supersymmetry
transformations are \bea \delta x^\mu=i\epsilon^\mu\xi^5,\quad
\delta\xi^\mu=\epsilon^\mu,\quad \delta\xi^5=\epsilon^5, \eea and
again the  lagrangian changes by a total derivative
\be\label{variation} \delta L^C=\frac{d}{d\tau}\left(-\beta\frac i
2\epsilon^\mu \xi_\mu-\gamma\frac i2\epsilon^5{\xi^5}\right). \ee
The generators of the global supersymmetries  are
\bea\label{susygen}
G_\mu&=&\pi_\mu+\beta\frac{i}2\xi_\mu+ip_\mu\xi^5=
i\beta\xi_\mu+ip_\mu\xi^5,
\nn\\
G_5&=&\pi_5+\gamma\frac{i}2\xi^5. \label{VsusyG}\eea They satisfy
\be \{G_\mu,G_\nu\}^*=-i\beta\eta_{\mu\nu},\quad
\{G_\mu,G_5\}^*=-ip_\mu,\quad \{G_5,G_5\}^*=-i\gamma. \ee At the
quantum level we have \be [G_\mu,G_\nu]_+=+\beta\eta_{\mu\nu},\quad
[G_\mu,G_5]_+=+p_\mu,\quad [G_5,G_5]_+^*=+\gamma.
\label{canoalgebra}\ee This is a canonical realization of the
starting algebra (\ref{eq:ext1}) and (\ref{eq:ext2}) with the
central charges\footnote{It is possible to show (see ref.
\cite{nuovo}) that the representations of the Vector Super
Poincar\'e algebra are characterized, besides the momentum square
and the Pauli-Lubanski invariant, by the quantity ${\sqrt{|Z\tilde
Z|}}$ and by the signs of $Z$ and $\tilde Z$.} \be Z=-\beta,\qquad
\tilde Z=-\gamma ,\qquad Z \tilde Z=-{\alp}^2. \ee

\section{Quantization in Reduced Space}

In this section we discuss the quantization of this system in terms
of the unconstrained variables. Let us  see the classical form of
the canonical action in the reduced space. The starting point is the
canonical lagrangian $L^C$ defined in eq. \bref{LagC}. It is locally
 supersymmetric and it is invariant  under
reparametrization in $\tau$. These two local symmetries are
generated by the first class constraints $\chi_5=0$ in
\bref{priconst} and $\phi=0$ in \bref{massshell} respectively.
 We  fix these gauge freedom by imposing the conditions
\be
x^0=\tau,\qquad \xi^5=0.
\ee
The first class constraints $\phi=\chi_5=0$ become
second class and are solved for $p_0$ and $\pi_5$ as
\be
p_0=\pm \sqrt{{\vec p}^{\,2}+{\alp}^2},\qquad \pi_5=ip_\mu\xi^\mu.
\ee
Other second class constraints are also used to reduce the variables
leaving  $x^i, p_i$ and $\xi^\mu$ as the independent variables  . 
The non-trivial Dirac brackets with respect to all these new second
class constraints are \be \{p_i,x^j\}^{**}=-{\delta_i}^j,\qquad
\{\xi^\mu,\xi^\nu\}^{**}=\frac{i}\beta{\eta^{\mu\nu}}.
\label{DB21}\ee The canonical form of the lagrangian \bref{LagC} in
the reduced space\footnote{ An analogous discussion for the
lagrangian of spinning particle of \cite{Brink:1976sz} was done in
reference \cite{Gauntlett:1990xq}.} becomes \be L^{C*}= \pm
\sqrt{{\vec p}^{\,2}+{\alp}^2} +\,\vec{p}\,\dot{\vec{x}}
-\beta\frac{i}2\xi_\mu\dot\xi^\mu.\label{lagcf} \ee

Now we quantize the model. The basic canonical (anti-)commutators
are \be [x^i, p_j]=i {\delta^i}_j,\qquad
[\xi^\mu,\xi^\nu]_+=-\frac{1}\beta{\eta^{\mu\nu}}.\label{qcomred}\ee
Note that  the sign degree of freedom  of $p_0$ must be taken  into
account in the quantum theory. The hamiltonian for the lagrangian
\bref{lagcf} is an operator \be P_0=
\pmatrix{{\omega}&0\cr0&-{\omega}},
\qquad {\omega}\equiv\sqrt{{\vec p}^{\,2}+{\alp}^2}
,\label{P0}\ee
taking eigenvalues $\pm {\omega}$ on the positive and negative energy eigenstates.
The Schr\"odinger equation becomes
\be i\pa_\tau \Psi(\vec x,\tau)=P_0 \Psi(\vec x,\tau),\qquad
\Psi=\pmatrix{\Psi_+\cr\Psi_-},
\label{Schgaugefixed}\ee
where $\Psi_+$ and $\Psi_-$ are 
positive and negative energy states. Now we look for a realization
of $\xi^\mu$. Since  they satisfy the anti-commutators in
\bref{qcomred} and must commute with all the bosonic variables, in
particular with the energy $P_0$ in \bref{P0}. We can realize them
in terms of 8-dimensional gamma matrices \be
\xi^\mu=\sqrt{\frac{-1}{2\beta}}\;\Gamma^\mu\Gamma^5=\sqrt{\frac{-1}{2\beta}}\;
\pmatrix{\gamma^\mu&0\cr0&-\gamma^\mu}\pmatrix{\gamma^5&0\cr0&-\gamma^5}
,\label{eq:5.8} \ee where the $\gamma^\mu$ and $\gamma^5$ are the
ordinary 4-component gamma matrices in 4-dimensions.
 where
$\Psi$ is an 8 component wave function and $\Psi_+$ and $\Psi_-$ are
four dimensional spinors associated to positive and negative energy
states. We rewrite the Schr\"odinger equation in a more familiar
{form} using a unitary transformation, an inverse Foldy-Whouthuysen
(FW) transformation, \be i\pa_\tau\tilde\Psi=
\pmatrix{{\balpha}^ip_i+\bbeta {\alp}&\cr&{\balpha}^ip_i-\bbeta
{\alp}} \tilde\Psi, \quad\Psi\equiv
S\pmatrix{U_4U_3^+U_4U_2U_1&\cr&U_4U_3^-U_4U_2U_1}\tilde\Psi,
\label{Schgaugefixed2}\ee where $\bbeta$ and ${\balpha}^i$ are usual
Dirac matrices and \bea
S=\pmatrix{1_2&&&\cr&&1_2&\cr&1_2&&\cr&&&1_2},\quad
U_3^+&=&\pmatrix{{e^{i\, {\frac{\theta_3}{2}\sigma_2}}}&\cr
&{e^{-i\, {\frac{\theta_3}{2}\sigma_2}}}},\quad
U_3^-=\pmatrix{{e^{i\, {\frac{\pi-\theta_3}{2}\sigma_2}}}&\cr
&{e^{-i\, {\frac{\pi-\theta_3}{2}\sigma_2}}}},
\nn\\
U_4=\pmatrix{1&&&\cr&&1&\cr&1&&\cr&&&1},\quad U_1&=&\pmatrix{{e^{i\, {\frac{\theta_1}{2}\sigma_3}}}&\cr&
{e^{i\, {\frac{\theta_1}{2}\sigma_3}}}},\qquad
U_2=\pmatrix{{e^{i\, {\frac{\theta_2}{2}\sigma_2}}}&\cr&
{e^{i\, {\frac{\theta_2}{2}\sigma_2}}}},\qquad
\nn\\
 \tan\theta_1&=&\frac{p_2}{p_1},\quad
\tan\theta_2=\frac{\sqrt{p_2^2+p_1^2}}{p_3},
\quad \tan\theta_3=\frac{\sqrt{\vec p^{\,2}}}{{\alp}}.\eea
 Multiplying by
$\Gamma^0=\pmatrix{\bbeta&\cr&-\bbeta}$  on the equation
\bref{Schgaugefixed2}, we get \be
 \Gamma^0i\pa_\tau\tilde\Psi= ({\Gamma^i(-i\pa_{x^i})+ {\alp}})\tilde\Psi,
\label{Schgaugefixed3}\ee where \be \gamma^i=\bbeta{\balpha}^i,\quad
\bbeta^2=1,\quad  \Gamma^\mu=\pmatrix{\gamma^\mu&\cr&- \gamma^\mu}.
\ee \bref{Schgaugefixed3} is the 8-components Dirac equation
reducible into two  4-components Dirac equations with mass $\mu$,

\be ({\Gamma^\mu(-i\pa_{x^\mu})+ {\alp}})\tilde\Psi(x^\mu)=
\pmatrix{-i\pa_{x^\mu}\gamma^\mu+ {\alp}&
    \cr& {}i\pa_{x^\mu}\gamma^\mu+ {\alp}}\tilde\Psi(x^\mu)=0 .
\label{Schgaugefixed4}\ee

\section{Quantization in Clifford Representation}

In this section 
the system is quantized in a covariant manner by requiring the
first class constraints to hold on the physical states.
The Dirac brackets are replaced by
the following graded-commutators, \bea
[p_\mu,x^\nu]=-i{\delta_\mu}^\nu,\qquad
[\xi^\mu,\xi^\nu]_+=-\dd{\frac 1 \beta{\eta^{\mu\nu}}},~~~
[\pi_5,\xi^5]_+=-i.
\label{ACM0A} \eea The odd variables define a Clifford algebra. This
is better seen by introducing a new set of variables with
appropriate normalization. We define \be
\lambda^\mu=\sqrt{-2\beta}\xi^\mu, ~~~\lambda^5= -i{\sqrt{\frac
2\gamma}}\left(\pi_5-\frac i 2\gamma\xi^5\right),~~~
\lambda^6=-i{\sqrt{\frac 2\gamma}}\left(\pi_5+\frac i
2\gamma\xi^5\right).\label{new_variables}\ee The reason to introduce
the $i$'s is that the momentum $\pi_5$ is anti-hermitian at the
pseudo-classical level and in this way  all the dynamical variables
are real, $(\lambda^A)^*=\lambda^A$.

The $\lambda^A$'s define a Clifford algebra $C_6$, \be
[\lambda^A,\lambda^B]_+=2{\ba\eta}^{AB},~~~{\ba\eta}^{AB}=(-,+,+,+,+,-),~~~
(A,B=0,1,2,3,5,6).\ee These variables can be identified as a
particular combination of the elements of another $C_6$ algebra
having within its  generators the $\Gamma^\mu$'s isomorphic to the
Dirac matrices  already used in eq. (\ref{eq:5.8}). This algebra is
defined by the following elements\bea \Gamma^\mu=\pmatrix{\gamma^\mu
&0\cr0&-\gamma^\mu},\qquad \Gamma^5=\pmatrix{\gamma^5
&0\cr0&-\gamma^5},\qquad \Gamma^6=\pmatrix{0&1\cr-1&0},
\label{Gamma6DrepA}\eea  satisfying   \be
[\Gamma^A,\Gamma^B]_+=2\tilde\eta^{AB},~~~~\tilde\eta^{AB}=(+,-,-,-,+,-).\ee
 Of course, both  Clifford algebras
have the same  automorphism group $SO(4,2)$. They are related in the
following way \bea
\lambda^A=\left\{\matrix{\Gamma^A\Gamma^5,&A=0,1,2,3,\cr
                  \Gamma^5,&A=5,\cr
                  i\Gamma^5\Gamma^6,&A=6.}\right.\eea

Let us start considering the $C_6$ generated by the $\lambda^A$'s.
The unitarity of the representation in terms of $\Gamma^A$'s
requires an extra measure, $\Gamma_*$, in the inner product, \be
<\Phi|\Psi>=\int d^4x\, {{ \Phi^\dagger}(x)}\Gamma_* {\Psi(x)}\equiv\int
d^4x\, {{\ba\Phi}(x)}{\Psi(x)}.
 \ee
In the quantization process we are going to require that the
operators in the matrix basis satisfy, with respect to the metric
$\Gamma_*$, the same reality property as in the classical case. We
find that the following operator satisfy our requirement
 \be
\Gamma_*=-i\lambda^0\lambda^6,~~~\Gamma_*^\dagger=\Gamma_*,~~~\Gamma_*^2=1,\ee
in fact \be{\ba\lambda}^A \equiv
\;\Gamma_*(\lambda^A)^\dagger\Gamma_*\;=\;\lambda^A. \ee
In terms of  $\Gamma^A$ we have
also \be \Gamma_*=\Gamma^0\Gamma^6.\ee

The expressions for the generators of  vector SUSY transformations
\bref{susygen} are obtained by inverting the relations
(\ref{new_variables}) \be \pi_5= \frac
i2\sqrt{\frac{\gamma}2}(\lambda^5+\lambda^6),~~~\xi^5=-\frac
1{\sqrt{2\gamma}} (\lambda^5-\lambda^6).\ee We find \bea
G^\mu=i\beta\xi^\mu+ip^\mu\xi^5&=&\frac
i{\sqrt{2\gamma}}\left({\alp}\lambda^\mu-p^\mu(\lambda^5-\lambda^6)\right)
\nn\\&=& \frac
i{\sqrt{2\gamma}}\Gamma^5\left({\alp}\Gamma^\mu-p^\mu(1-i\Gamma^6)\right)\eea
and \be G_5=\pi_5+i\frac\gamma 2 \xi^5=i\sqrt{\frac\gamma
2}\lambda^6=- \sqrt{\frac\gamma 2}\Gamma^5\Gamma^6.\ee The
generators have the following conjugation properties \be {\ba
G_\mu}=-G_\mu,~~~{\ba G}_5=-G_5.\ee In analogous way we get the
expression for the odd first class constraint \be
\chi_5=\pi_5-i\frac{\gamma}2\xi^5-ip_\mu\xi^\mu=\frac
i{\sqrt{-2\beta}}\left({\alp}\lambda^5-p_\mu\lambda^\mu\right)=\frac
i{\sqrt{-2\beta}}\Gamma^5(p_\mu\Gamma^\mu+{\alp}).\ee The
requirement that the first class constraint $\chi_5$  holds on the
physical states is equivalent to require the Dirac equation on an
8-dimensional spinor $\Psi$ \be (p_\mu\Gamma^\mu+{\alp})\Psi=0.\ee
The other first class constraint \bref{massshell},
$\phi=\frac12(p^2+{\alp}^2)=0$, is then automatically satisfied
since, \be \phi\;\Psi=\frac12\;  {\cal D}^2\,\Psi=0,\quad {\rm
with}\quad
 {\cal D}\equiv\Gamma^5(p_\mu\Gamma^\mu+{\alp}).\ee
At the pseudo-classical level the first class constraints are
invariant under the supersymmetry transformation. At the quantum
level this is reflected by the following properties \be [G_\mu,
{\cal D}]_+=[G_5, {\cal D}]_+=0.\ee From these relations we can
define the corresponding symmetry transformations on the wave
function if we can construct a matrix, call it $E$, anticommuting
with all the dynamical variables, $\lambda^A$'s. In this case we
have \be [EG_\mu, {\cal D}]=[EG_5, {\cal D}]=0\ee and  the
transformations generated by $EG_\mu$ and $EG_5$ leave invariant the
action of the theory \be\int d^4x\;\bar\psi  {\cal D}\psi.\ee The
corresponding  unitary transformations (with respect to the metric
$\Gamma_*$) are \be e^{i\alpha_5 EG_5},~~~~e^{i\alpha^\mu
EG_\mu},\ee where $\alpha_5$ and $\alpha^\mu$ are even parameters
defining the transformations\footnote{ Note that ${\ba{ EG_5}}=
EG_5,~~~{\ba{ EG_\mu}}= EG_\mu$}.  The operator $E$ can be easily
constructed since our Clifford algebra is defined in a even
dimensional space. Therefore, \be
E=\lambda^7=i\lambda^0\lambda^1\lambda^2\lambda^3\lambda^5\lambda^6=
-\Gamma^0\Gamma^1\Gamma^2\Gamma^3\Gamma^6=i\Gamma^5\Gamma^7, 
\qquad \Gamma^7\equiv
i\Gamma^0\Gamma^1\Gamma^2\Gamma^3\Gamma^5\Gamma^6,\ee anti-commutes
with all $\lambda^A$'s. One could ask if it is possible to recover
the result of ref. \cite{Brink:1976sz}, that is a Dirac equation in
a 4-dimensional spinor space. This was indeed done in ref.
\cite{Barducci:1976qu} where it was imposed a further constraint
\be\pi_5+i\frac{\gamma}2\xi^5=0.\ee In this way the quantization can
be done by using only a $C_5$ algebra which can be realized in a
4-dimensional space. Note that if we impose this condition the
supersymmetry generator $G_5$ vanishes identically. Therefore one
looses the rigid supersymmetry although the local one remains.

\section{BPS Configurations}

Here we will consider the BPS equations for the massive spinning
particle. The corresponding  bosonic supersymmetric configurations
appear only when the lagrangians have a gauge world-line
supersymmetry.

The lagrangian of the massive spinning particle \bref{actionv} has a
gauge symmetry when the parameters ${\alp},\beta,\gamma$  verify the
condition \bref{BPScond},
\be\label{condition12}{-\beta\gamma}={{\alp}^2}
.\ee Now we look for supersymmetric bosonic configurations. For
consistency we look for transformations of the fermionic variables
not changing their initial value that is supposed to vanish \bea
0=\left.\delta\xi^5\right|_{\rm fermions=0}&=& \epsilon^5+
\kappa^5,\nn\\
\nn 0=\left.\delta\xi^0\right|_{\rm fermions=0}&=&
\epsilon^0-\frac{{\alp}}{\beta}\cosh v\;\kappa^5~~~
\\
 0=\left.\delta\xi^i\right|_{\rm fermions=0}&=&
\epsilon^i-\frac{{\alp}}{\beta}\frac{v^i}v\sinh v \;\kappa^5.~~~
\eea The previous equations have a non-trivial solution
 if \be v^i={\rm constant}. \label{BPS}\ee Then, all the parameters
 can be expressed in terms of an independent global
supersymmetry parameter, $\epsilon^5$, as \be
\kappa^5=-\epsilon^5,\qquad \epsilon^0=-\frac{{\alp}}{\beta}\cosh
v\;\epsilon^5 ,\qquad
\epsilon^i=-\frac{{\alp}}{\beta}\frac{v^i}v\sinh v \;\epsilon^5. \ee
Equation \bref{BPS} is the BPS equation of this model. If we write
this expression in terms of space-time coordinates, using the solutions
(\ref{eq:3.15}) of \bref{eq69a},  we get \be \frac{{\dot
x}^\mu}{\sqrt{-{\dot x}^2}}= constant.\label{BPS2}\ee In Hamiltonian
terms this implies that the momentum is constant. Note that this BPS
equation implies the second order equations of motion of a free
relativistic particle. Therefore the BPS configurations  \bref{BPS2}
preserve 1/5 of the supersymmetry. Notice that the fraction of
 preserved supersymmetry is different from the ordinary (spinor realization of)
Super Poincar\'e group as, for example, in the case of the
superparticle.

\section{Discussions}

In this paper we use the rigid space-time vector supersymmetry to
construct the action of the massive spinning particle from the
non-linear realization method.

For particular values of the coefficients of the lagrangian, the
model has world line gauge supersymmetry which is the analogous of
the fermionic kappa symmetry of the superparticle case.  By
quantizing the model in such a way to respect the rigid
supersymmetry, we find two decoupled 4d Dirac equations with the
same mass. The supersymmetry transformations at quantum level mix
the two 4d Dirac equations.

At classical level we find BPS configurations that preserve 1/5 of
the supersymmetry. The BPS equations imply  second order equations
of motion.

In a future work\cite{nuovo} we will study the representations of
the Vector Super Poincar\'e algebra which, in the massive case, are
characterized by the quantity ${\sqrt{|Z\tilde Z|}}$ and by the
signs of $Z$ and $\tilde Z$. The massless spinning particle will be
also considered.

Two interesting questions are:  the possible physical role of the
space-time vector supersymmetry in quantum field theories and the
relation of this approach with the one based on space-time spinorial
supersymmetry.

\vskip 3mm

{\bf Note added} After this paper was put on the archive. M.
Plyushchay has informed of a previous work \cite{Gamboa:1997fs}
which has some overlap with our work. In particular about the
constraints analysis of the model we have considered and the
non-equivalence among the models of references
\cite{Barducci:1976qu} and \cite{Brink:1976sz}.

\vspace{1cm}

 {\bf Acknowledgements } We are grateful to Luis Alvarez-Gaume, Massimo Bianchi, Dan
Freedman, Gary Gibbons, Jaume Gomis, Mikhail  Plyushchay Paul
Townsend and Toine Van Proeyen for interesting discussions. This
work has been supported by the European EC-RTN project
MRTN-CT-2004-005104, MCYT FPA 2007-66665 and CIRIT GC 2005SGR-00564.
One of us, J.G. would like to thank the Galileo Galilei Institute
for Theoretical Physics for its hospitality and INFN for partial
support during part of the elaboration of this work.


\begin{thebibliography}{37}
\bibitem{nuovo}
The representations of this extension of the Poincar\`e algebra will
be studied in a future paper.
\end{thebibliography}
\end{document}